\documentclass[12pt,preprint,longnamesfirst]{aastex}
\usepackage{emulateapj5}
\usepackage{graphicx}

\slugcomment{Submitted to Astrophysical Journal Letters}
\shorttitle{SN~1993J Light Echoes}
\shortauthors{Sugerman \& Crotts}

\begin{document}

\title{Multiple Light Echoes from Supernova 1993J}

\author{Ben E.K.~Sugerman and
Arlin P.S.~Crotts}
\affil{Dept.~of Astronomy, Columbia University, 550 W.~120th St., New
York, NY 10027 U.S.A.}
\email{ben@astro.columbia.edu, arlin@astro.columbia.edu}

\begin{abstract}
Using the technique of PSF-matched image subtraction, we have analyzed
archival {\em HST}/WFPC2 
data to reveal details of at least two light-echo structures,
including some unknown before now, around SN~1993J in the galaxy M81.
In particular, we see one partial sheet of material 81~pc in front of the SN
and tilted $\sim60^\circ$ relative to the disk plane of M81, and another
220~pc in front of the SN, roughly parallel to the disk.
The inferred echoing material is consistent with the \ion{H}{1}
surface density detected in this region of M81's disk; however, these
data imply a fragmented covering factor for the
echoing structures. We discuss prospects for future (roughly annual)
visits by {\em HST} to image these and yet undiscovered echoes in the
interstellar and circumstellar environment of SN~1993J. 
\end{abstract}


\section{Introduction}

Supernova (SN) 1993J was a type II SN 
\citep{rip93} in
the nearby galaxy NGC 3031 (M81), the closest SN seen in the past
decade. It's spectrum quickly lost most of its hydrogen emission
lines, indicating that most of the progenitor's hydrogen envelope had
been stripped, and justifying classification of the SN as type IIb
\citep{nom93}. The lost material probably formed a circumstellar
shell, as evidenced by emission in X-rays \citep{zim94}, radio
\citep{bar94}, and narrow optical lines \citep{bnt94,mat00}.

Since the SN exploded near a spiral arm in M81, it is expected to
illuminate interstellar and circumstellar material in the form
of light echoes. Such echoes have been reported for e.g.,\
SN~1998bu \citep{cap01}, SN~1991T \citep{sch94,spa99} and SN~1987A
\citep{cro88a}. 
Light echoes from SN~1987A have revealed structures on interstellar
and circumstellar scales \citep[and references therein]{cro95,cro01}, which
have been used to map the surrounding material in three dimensions and
tie it to kinematic information \citep{xu95,xu99,cro00}, offering
unique insights into the history of the associated stars and gas. 

Recently, \citet{liu02} discovered a light echo around SN~1993J
based on data taken on 2001 June 4 on the WF4 detector of the Wide
Field Planetary Camera 2 (WFPC2) aboard the {\em Hubble Space
Telescope} ({\em HST}).
We have reanalyzed these now publicly available data, and others (\S2),
and have found yet another distinct echo. Using image subtraction and
PSF-fitting techniques we have also produced detailed analyses
(\S3) of the structure and reflectivity of both echoing
structures. These results already reveal intriguing details about the
interstellar medium in M81, and are likely to continue to do so in the
future.   

\section{Observations and Reductions}

The SN has been visited with {\em HST}/WFPC2 several times: UT 1994
April 18, 1995 January 31 and 2001 June 4, respectively 1.00, 1.79 and
8.13y after maximum light [UT 19 April 1993; Benson et al.~(1994)].
We make use of publically-available data listed in Table \ref{tbl1};
in other bands and at other epochs, the data are insufficient in
exposure time or number of visits for useful comparison.

Pipeline-calibrated images were processed as in \citet{Sug02}.  When
necessary, stars were removed with Tiny Tim model PSFs.  All images
were geometrically registered to a common orientation with residuals
$\le 0.1$ pixel rms.  Light echoes are transient sources, and are best
detected via {\em difimphot} image-subtraction \citep{TC96}, in which
Fourier techniques match empirically-derived stellar profiles between
images to remove sources of constant flux.  Even with PSF-matching,
reduced data can be unusually noisy -- we could not difference the
1995 F439W from the 2001 F450W images, since the noise from resampling
PC images to the highly-undersampled WF4 resolution, and from the
different passbands, was greater than any echo signal.  Instead,
stellar sources were removed from the 2001 F450W integration using
{\em daophot}.  We use standard {\em HST} photometric calibrations,
which we checked against secondary standards used to monitor SN~1993J
\citep[L. Wells, private communication;][]{sil94}, and which agree to
within 5\%.  We measured the SN centroid in the 1994 epoch
(with $V \approx 19$), providing its unambiguous position.

\section{Analysis and Discussion}
 
Figure 1a shows the F555W image from 2001, and Figure 1b shows its
difference from the F555W integration of 1995.  The 450W
image from 2001, with stellar sources removed (\S2),
is shown in Figure 1c, and the difference image in F814W (with
obviously lower signal-to-noise ratio) corresponding to panel (b) is
show in Figure 1d.  These clearly show two echoes from 2001 (lighter
shade) as well as a confusion due to poor subtraction within $\sim
0\farcs4$ of the SN.  The outermost echo is seen at radii extending
from $\theta=1\farcs84$ to $1\farcs95$ from the SN, at position angles
170$<$PA$<$290, with the largest radii near PA 225.  These correspond
to distances from the SN sightline $r = \theta D = 32.4$ to 34.3~pc
($D = 3.63$~Mpc: \citealt{fre94}).  Since it is a light echo, one can
compute the foreground distance (along the SN sightline) $z = r^2/2ct
- ct/2 = 209$ to 235 pc, implying a tilt of about 37$^\circ$ with the
southwest side farther in front of the SN. An echo at the same $z$
distance would occur at $\theta=0\farcs89$ on 1995 January 31, and
indeed there is some marginal indication of such a feature (dark in
Figures 1b and 1d) between PA 190 and 260.  This is far from definite,
but may indicate a shrinking of the echo cloud in PA at earlier
epochs, hence the possibility that the echo cloud does not extend in
front of the SN.

The inner echo lies at $\theta\sim1\farcs15$ (all at the same radius,
to within the errors) in 2001, over 0$<$PA$<$60.  In 1995, this same
$z$ would correspond to $\theta \approx 0\farcs55$, which is
bordering the confusion region of the bright PSF of the SN.  The
echo lies at a foreground distance $z = 81$~pc, and is perpendicular
to the line of sight to within about 25$^\circ$.  The geometry of all
echoes is shown in Figure 2. Radial profiles through the echoes in
F555W are plotted in Figure 3, demonstrating that these detections are
above the background noise (e.g. $\sigma_{\rm F555W}=1.5$~DN~pix$^{-1}$).  

Using the naming convention of \citet{xu95}, we denote the outer echo as
SW770 and the inner echo as NE260.  M81, with an angular momentum
vector inclined 59$^\circ$ along PA 62 \citep{rot75} (such
that the southwest side of the disk is closer to Earth), and SN~1993J,
southwest of the nucleus of M81, would imply that the tilt of the
NE260 dust sheet is roughly perpendicular to the disk of the galaxy.
In comparison, the SW770 echo is inclined to within $\sim30^\circ$ of
the disk plane.  The SN, with its massive progenitor
contained within a dense gas cloud, is presumed to lie near the disk plane
(shown in Fig.\ 2).  We thus detect two dust structures,
both extending more than one gas scale-height (c.f.\ \citealt{bro98})
above the plane.  SW770 sits a roughly constant 110 pc above the disk
plane in M81, while NE260 appears to miss this plane by $\ga$40~pc and
extends at least 60 pc above the disk.  Without requiring the SN
to lie in the disk plane, one might hypothesize that the
disk plane passes near both echoing structures, implying the SN may
lie $\sim70-90$~pc behind the disk.

%
SW770 has measurable surface brightnesses over the PA range
160--280. Over 190$<$PA$<$250, all three bands (F450W, F555W and
F814W) track each other in surface brightness consistently.  For
$190<$PA$<250$, $\langle \mu \rangle = 23.3\pm0.1$, $23.3\pm0.1$ and
$24.6\pm0.2$ in STMAG, respectively transforming to vegamag colors\footnote{
Color transformations from STMAG to Johnson-Cousins (vegamags) were 
determined using {\em synphot} and the Bruzual-Persson-Gunn-Stryker
Spectrophotometry Atlas.} of roughly
$B-V=0.6\pm0.2$ and $V-I_C=0.0 \pm 0.3$.  The spatial variation in
surface brightness is similar in $B$ and $V$, both rising with
inreasing PA values.  At PA$\approx270$, $\mu_{450}=23.0\pm 0.1$ and
$\mu_{555}=23.1 \pm 0.1$, while at PA$\approx$180, $\mu_{450}=23.6 \pm
0.2$ and $\mu_{555}=24.1 \pm 0.2$.  In contrast $\mu_{814}$ is nearly
equally faint at both extremal PAs, $\mu_{814}=25.1 \pm 0.4$, implying
colors at PA$\sim270$ of
$B-V=0.5 \pm 0.3$ and $V-I_C=-0.9 \pm 0.5$.
In comparison, NE260 has
approximately the same global colors as SW770 (over 190$<$PA$<$250):
$B-V=0.5 \pm 0.4$ and $V-I_C=0.2 \pm 0.4$, and there is little
evidence for such a color gradient.  The surface brightnesses
themselves are fainter by about 0.5 mag arcsec$^{-2}$ compared to
SW770.

In order to interpret these surface colors in terms of reflectivity,
we must know the colors of the incident echoing flux.  Integrating
over the entire SN lightcurve \citep{ben94,ric94} from $\sim3$ to 127
days after core collapse yields $B-V=0.73$, $V-I_C=0.69$ for the fluence
of the (nearly) entire event. 
The color 
change due to dust reflectivity for NE260 and SW770 (with PA$<$250)
is $\Delta (B-V)=-0.1$ and $\Delta (V-I_C)=-0.7$, while for SW770 with
PA$>$250, $\Delta (B-V)=-0.2$ and $\Delta (V-I_C)=-1.6$.  
The largest changes in color are imparted in the 
Rayleighan scattering regime by very small particles.
Integration of the scattering efficiency $S(\lambda,a)$
\citep{xu94} using the
dust scattering parameters of  \citet[and references therein]{wd01},
we find $S \propto \lambda^{-4.3}$ for $a<0.01 \mu m$, yielding
$\Delta (B-V)_{\rm max}=-0.96$ and $\Delta (V-I_C)_{\rm max}=-1.72$.  
As the observed color shifts are smaller, 
the echoes should be consistent with a galactic dust distribution.
Dust modeling will be examined in detail by \citet{Sug03}.

Attributing the blue color of SW770 at PA$>$250 to a small-grain-only
hypothesis is sufficiently improbable as to warrant alternative
explanations.  Any extinction mechanism for F814W light should
more-strongly affect the F450W and F555W bands.  One might invoke
additional flux from mechanisms beyond direct reflection to increase
the flux in F450W and F555W over F814W for PA$<$250.  Extended red
emission \citep{wit90a} would contribute significantly to the F814W
band but not the bluer bands.  However, we caution that this very blue
$V-I_C$ SW770 color is near the detection threshold, and we cannot
rule out a ``hot pixel'' or variable star producing an erroneous
image-subtraction residual in the F814W image.
 
For the sake of this discussion, we adopt a dust model with
isotropically-scattering grains.
We note that this assumption disagrees with some measurements
of Galactic interstellar dust \citep{wit90b,mat79,tol81}.  Using
this model, we calculate the ratio of dust densities in the two
echoes from the surface brightnesses and echo geometry.
For an echo cloud that is thick relative to the the depth of dust
echoing at a given time ($t_{s} \bigl | {{dz} \over {dt}} \bigr |$),
the surface brightness is predicted by \cite{che86}:
$$\mu (\theta) = {{n_{d} Q_{s} \sigma_{d} } \over {4 \pi D^{2}}} ~
{{L_{\nu} t_{s}} \over {4 \pi R^{2}}} ~ \left| {{dz} \over {dt}}
 \right| ~
F(\alpha),$$

\noindent where the average apparent luminosity $L_{\nu}/4 \pi D^{2}$
over the SN light pulse duration $t_{s}$ is observed directly from the
SN.  The geometric factors $R = \sqrt{(z^2 + r^2)}$ (the SN-to-cloud
distance), $\alpha$ (the scattering angle), and $|dz/dt| = r^2/2ct^2 +
c/2$ (describing the depth of the echoing region), are determined
precisely by $\theta$ from the light-travel-delay equation of an echo.
This leaves: the grain scattering efficiency $Q_s$ and geometric
cross-section $\sigma_d$ (assumed equal in the two echoing
clouds); the dust number density $n_{d}$; and the scattering phase
function $F(\alpha)= (1-g^2) / (1+g^2-2g \cos{\alpha})^{3/2}$ [Henyey
\& Greenstein 1941 -- which does quite well for small $\alpha$ such as
here \citep{wit89}], where $g=\overline {\cos{\alpha}}$ is the degree of
forward scattering, here assumed to be zero.
The ratio of the geometric factors $F(\alpha)|\dot{z}|/R^2$ for the
two echoes is 1.52
(inner/outer), but SW770 is spread over $r_{outer}/r_{inner} = 1.71$
times the area.  Since the echoes are underresolved by WFPC2, SW770
should be about 1.13 times higher observed surface brightness than
NE260, if they have the same dust properties and density.  In truth,
we measure SW770 to be about 1.6 times brighter, implying either a 40\%
higher dust density in SW770, or the SW770
cloud is 40\% thicker along the line of sight.  If, instead, the
clouds are geometrically thin compared to 
$t_{s} \left|\frac{dz}{dt} \right|$, 
geometric factors would predict an inner echo
brightness 2.56 times that of SW770 (for equivalent dust), implying
that the SW770 cloud contains 4.1 times higher dust surface density.
How do these numbers change if we instead invoke a non-zero $g$?
For reasonable values ($g \approx 0.5$), the above brightness ratios
change by only 1\% or less, since the scattering angles for the two
echoes are very similar ($\alpha = 10\fdg5$ for NE260 versus $8\fdg5$
for SW770.)

The echo's effective width is given by $w = t_s ~ dr/dt$ where $dr/dt =
c(z+ct/2)/\sqrt{(2z+ct/2)ct}$, and $t_s$ is the total
fluence in $V$ ($6.31 \times 10^{-7}$ ergs cm$^{-2}$ \AA$^{-1}$)
divided by the maximum light flux, yielding $t_s = 3.5 \times
10^6$s and $w = 0.23$~pc. A thin sheet of
isotropic reflectors at the position of SW770 in year 2001 (with
170$<$PA$<$290) diverts no more than 0.0026\% of the SN flux seen at 
Earth.
%
%
We observe a flux in F555W of $\sim4.3\times 10^{-18}$
ergs cm$^{-2}$ s$^{-1}$ \AA$^{-1}$ from the echo, versus a corresponding
maximum flux from the SN of $1.8\times 10^{-13}$ ergs cm$^{-2}$ s$^{-1}$
\AA$^{-1}$, or about 0.0024\%.  SW770, over the position observed,
appears to be optically thick.  This implies NE260 also has $A_v \ga
\frac{1}{4}$.

For a dust albedo of 0.5 and grain diameter of 0.1~$\mu$m (the largest
Rayleigh-like particle), unit
optical depth corresponds to $\sim 15 ~\mu$g~cm$^{-2}$ for grains
of density 1~g~cm$^{-3}$.  If the gas-to-dust ratio is 100,
this corresponds to $N_H \approx 8\times 10^{20}$~cm$^{-2}$ for SW770.
At the position of the SN, $N_{\mbox{\footnotesize{H{\sc i}}}} \approx
10^{21}$~cm$^{-2}$, so SW770 structure is consistent with the dominant
locus of gas along the Earth-SN sightline.

The velocity structure of this region of the galaxy has been studied
in \ion{H}{1} 21~cm emission and optical/UV absorption (of the SN
itself).  This structure is relatively smooth and locally centered
near $v_{lsr} = -135$ km s$^{-1}$ \citep{rot75}, with some gas over
the range $-155 < v_{lsr} < -115$ km s$^{-1}$ \citep{ros75}.  In
absorption against SN~1993J, the predominant M81 interstellar
components are $-$119 and $-$135~km~s$^{-1}$, with possible lesser
components at $-110$ and $-100$~km~s$^{-1}$ \citep{vla94}.
The former two interstellar components are each at least twice as
strong in \ion{Ca}{2} column density as the latter two (hence
containing about 56\% and 26\% of the interstellar gas), and appear to
be cold.  In IUE spectra of UV absorption lines \citep{mar00}, a
strong component at $-$130~km~s$^{-1}$ and a weaker one at
$-$90~km~s$^{-1}$ is seen in low-ionization species, probably
consistent with the \ion{Ca}{2} components.

It is possible that the inner and outer echoes correspond to the two
dominant absorption features ($-$119 and $-$135~km~s$^{-1}$), but this
is difficult to state with certainty given the limited amount of data
and the partial covering factor of the structures involved.  The
structure and strength of the echoes seems to imply, however, that
major portions of interstellar material in this part of the disk may
be broken into fragments and perhaps even propelled a scale height or
more out of the disk plane.

\section{Future Prospects}

One prospect that is unlikely is using these echoes to measure the
distance to M81.  The maximum-polarization technique \citep{spa96}
requires 90$^\circ$ scattering, a condition unlikely for
these echoes, but potentially attainable for circumstellar echoes yet
to be observed.  (This has been implemented for SN~1987A's
circumstellar echoes, for instance - Crotts et al.~in preparation.)~
The use of {\em interstellar} echoes to measure the distance to M81 will
require centuries of observation before their power-law expansion
behavior begins to break down, and even then the intrinsic geometry of
the echoing structure seems unlikely to cooperate in performing a
distance determination.

As time progresses, we expect more light echoes will 
appear.  At small radii, one must compete with the bright
central source, making echo detection difficult for $\theta < 0.5$
arcsec, or $r < 8$~pc.  In year 2002 this radius
corresponds to $z \approx 16$ pc and will continue to decrease
(roughly as $1/t$).  This foreground distance is
well beyond the likely circumstellar region around the SN progenitor.
Eventually, the echo from
the circumstellar material itself might become apparent.  The SN
progenitor evidently has been emitting a dense wind with outflow
velocity of at least $\sim 10$~km~s$^{-1}$ \citep{fra96}, probably for
$\sim10^6$~y.  Depending on the density of interstellar material into
which this wind is propagating, dense circumstellar
nebulosity may extend beyond the PSF of the central source, and more
careful analysis 
may reveal echoes at even smaller radii.

Pursuing these and further echoes around SN~1993J with a series of PC (or ACS)
images would be valuable, particularly in bands near F555W, offering the best
sensitivity.  Since SW770 is
expanding at a rate corresponding to the PSF width of {\em HST} every
0.8~y, an annual visit to image the echoes of SN~1993J will
probe of new material each time.  Over the course of a decade or so,
we should be able to build a more detailed, three-dimensional view of
the interstellar medium in a $\sim10^6$~pc$^3$ volume of M81's spiral
arm, and begin to glimpse the outer edges of the region affected by
the mass loss from SN~1993J's progenitor.

\acknowledgments This research was supported by grants AST 02-06048
from the NSF and \#8806, 8872 and 9328 from the STScI.

\begin{deluxetable}{lccccc}
\tablecaption{WFPC2 data used in this work \label{tbl1}}
\tablewidth{0pt}
\tablehead{
 \colhead{Epoch} & \colhead{Detector} & \colhead{Filter} & 
 \colhead{$\lambda_c$\tablenotemark{a}} &
 \colhead{$t_{exp}$\tablenotemark{b}}  \\
 \colhead{} & \colhead{} & \colhead{} &
 \colhead{\AA} &  \colhead{sec} &
}
\startdata
1994 April 18   & PC & F555W & 5407 & 300   \\
1995 January 31 & PC & F439W & 4300 & 1200  \\
                & PC & F555W & 5407 & 900   \\
                & PC & F814W & 7940 & 900  \\
2001 June 4     & WF4 & F450W & 4520 & 2000 \\
                & WF4 & F555W & 5407 & 2000 \\
                & WF4 & F814W & 7940 & 2000 \\
\enddata
\tablenotetext{a}{Central Wavelength}
\tablenotetext{b}{Total exposure time}
\end{deluxetable}

\begin{figure}
\plotone{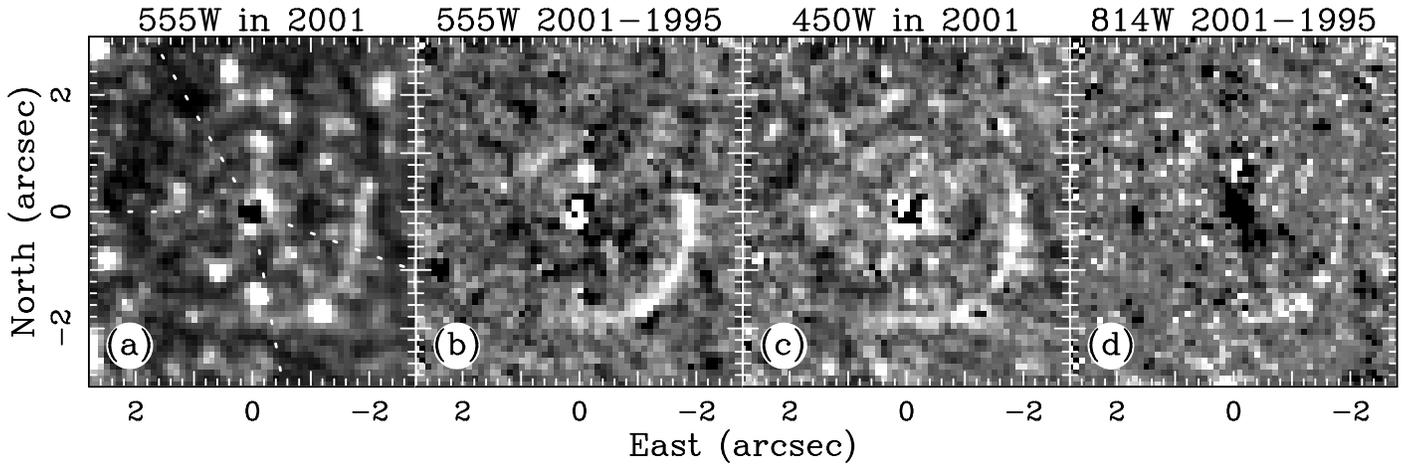}
\caption{ {\em HST} WF4 images of the region surrounding SN~1993J.  In
all panels, we have subtracted the SN and an adjacent star (PA=347,
$r=0\farcs72$), which we found to be variable.  (a) A direct image in
F555W from 2001; (b) a difference of F555W images between 2001 and
1995; (c) direct image in F450W from 2001, with all stellar sources
removed; (d) same as (b) but in F814W (image is scaled to enhance
faint pixels).  SW770 is visible between PA
160--280.  The fainter, inner echo NE260 is apparent from PA 10--60 at a
radius of $\sim1\arcsec$ in (b) and (c).  A faint negative region
around PA 190--260, $r=0\farcs85$ appears in panels (b) and (d),
suggestive of an echo in the 1995 integration.  Dotted white lines in
panel (a) show the locations of radial profiles from figure 3.}
\end{figure}

\begin{figure}
\epsscale{0.5}
\plotone{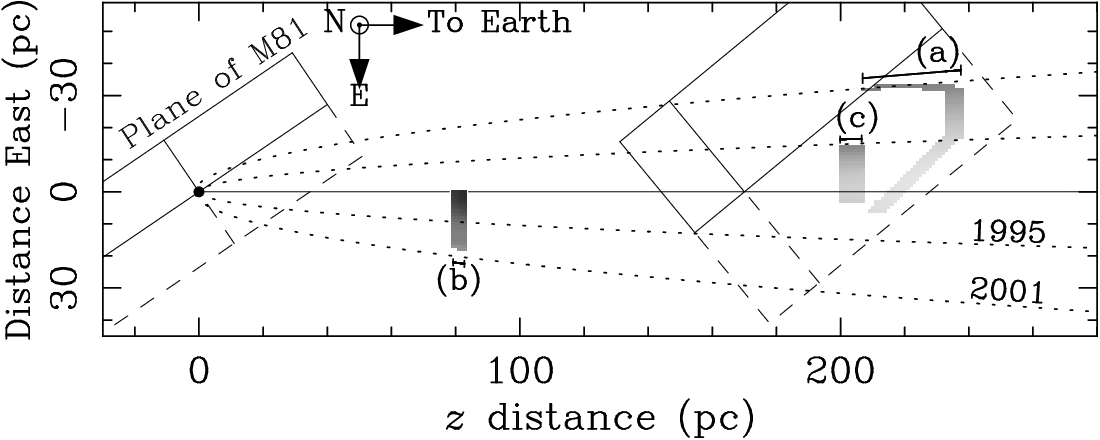}
\caption{Geometry of M81 and the light echoes.  Images in Figure 1
show the projected views of an observer at large $z$ distance. This
diagram shows the geometry from far to the north, with the plane of
the page separating north from south.  SN~1993J is at the origin.  The
orientation of M81's disk (see text) and the plane containing SW770
are indicated.  Echo parabolae from 1995 and 2001 are drawn with
dotted lines.  Echoes are marked as follows: (a) SW770 in 2001, (b)
NE260 in 2001, and (c) SW770 in 1995.  Grayscale indicates height of
the echoing material above (below) the page, with darker (lighter)
shades indicating greater northern (southern) position.  }
\end{figure}

\begin{figure}
\epsscale{0.35} \centering
\includegraphics[angle=-90,scale=0.35]{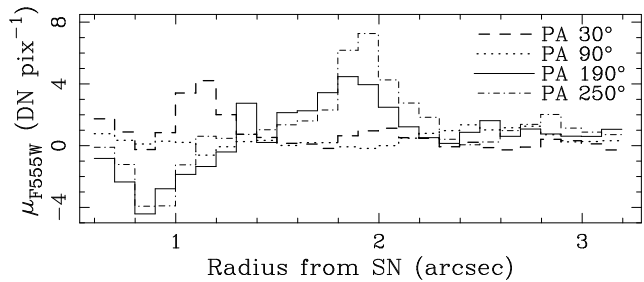}
\caption{Radial profiles of F555W surface brightness
  measured in 20\degr-wide annular bins centered on the PAs as marked
  here and in figure 1a.
  NE260 in 2001 is clearly visible above the noise at PA 30\degr, as
  are SW770 in 2001 and 1995 at PAs 190\degr\ and 250\degr.  For
  comparison, no such structure is evident at PA 90\degr\ or at large
  radii.}
\end{figure}

\end{document}